\documentstyle[aps,prb,twocolumn,epsfig,floats]{revtex}
\begin{document}
\wideabs{
\author{N.J. Clayton$^1$, H. Ito$^2$, S.M. Hayden$^1$, P. J. Meeson$^1$,
M. Springfield$^1$, G. Saint$^3$}
\address{
$^1$H.H. Wills Physics Laboratory, University of Bristol, Tyndall
Avenue, Bristol, BS8 1TL, U.K.
$^2$Department of Applied Physics, Nagoya University, Nagoya 464-8603, Japan\\
$^3$Department of Chemistry, Kyoto University, Kyoto 606-8502,
Japan }
\title{Superconducting Fluctuations and the Reduced Dimensionality
of the Organic Superconductor $\kappa$-
(BEDT-TTF)$_{2}$Cu(NCS)$_{2}$ as Observed through Measurements of
the de Haas-van Alphen Effect}
\date{5 July 2001}
\maketitle
\begin{abstract}
We report measurements of the de Haas-van Alphen (DHVA) effect and
AC susceptibility in the layered organic superconductor
$\kappa$-(BEDT-TTF)$_{2}$Cu(NCS)$_{2}$.  The amplitude of the DHVA
oscillations is attenuated over a wide field range above the
irreversibility line, $B_{\text{irr}}(B)$, below which a rigid
flux lattice is formed.  Thus the DHVA effect provides a unique
probe of the superconducting fluctuations at high fields in this
material.  We compare our measurements with other determinations of
the superconducting phase diagram of
$\kappa$-(BEDT-TTF)$_{2}$Cu(NCS)$_{2}$.
\end{abstract}
\pacs{74.70.Kn, 71.18, 74.25.Dw} }

\narrowtext

\section{Introduction}
$\kappa$-(BEDT-TTF)$_{2}$Cu(NCS)$_{2}$ is an organic
superconductor consisting of alternating layers of conducting
BEDT-TTF and insulating Cu(NCS)$_{2}$ anions \cite{Saito}.  This
complicated structure yields a simple Fermi surface consisting of
quasi two-dimensional (Q2D) pockets, and open quasi
one-dimensional (Q1D) sheets \cite{Oshima}.   The layered
structure is reflected in the anisotropic properties of the
superconductivity (T$_{c} \sim$ 9 K), appearing in transport
\cite{ito93,friemelfluc,friemelangle,brooks}, magnetization
\cite{ito,lang}, AC susceptibility \cite{Kanoda1271,Mansky}, and
torque magnetometry \cite{Farrell,Kawamata,sasaki,Mola2}
experiments.  The low dimensionality together with short coherence
length, result in fluctuations of the superconducting order
parameter being important over a wide range in temperature and
applied field. The importance of thermal
\cite{friemelfluc,ito,lang,Graebner,Behnia} or quantum
fluctuations \cite{sasaki,Mola2,Mota} leads to, for example, a
broadened anomaly in the specific heat capacity. The
superconducting phase diagram of $\kappa$-(BEDT-
TTF)$_{2}$Cu(NCS)$_{2}$ has a complicated structure
\cite{Mola2,Nishizaki,Pasquire}.   A vortex liquid phase is
thought to exist over a wide $B-T$ region bounded by the
irreversibility line at low temperatures and fields.

Recent experience has shown that the de Haas-van Alphen (DHVA)
effect is a powerful probe for in the investigation of the mixed
state of superconductors.  The DHVA effect consists of an
oscillatory variation of the magnetization $\tilde{M}$ which
varies periodically in inverse applied field.  For a
two-dimensional metal\cite{Shoenberg,Harrison,Champel2001}, each
Fermi surface sheet with area $A$, gives rise to a fundamental
oscillatory magnetization,
\begin{eqnarray}
\label{DHVA_eqn} \tilde{M}&=& -\frac{e^2 V F}{b m_e
\pi}\frac{X}{\sinh X} \exp(-\pi m_b / e B \tau_0) \nonumber \\
& & \times R_{s}(B) \times \cos \left( - \frac{2 \pi F}{B} +
\phi\right) ,
\end{eqnarray}
where $B$ is the applied field, $b$ is the thickness of the 2D
layer, $V$ is the volume of the sample, $F=(\hbar/2 \pi e)A$,
$m^{\ast}$ is the renormalised cyclotron mass, $m_b$ the band mass
and $X=2 \pi m^{\ast} k_{\text{B}} T /e \hbar B$. DHVA
oscillations in the superconducting mixed state were first
observed in NbSe$_2$ \cite{GR}, and have now been measured in many
superconductors, such as $A$15 materials \cite{JT}, borocarbides
\cite{Terashima}, organic superconductors
\cite{vanderWel,sasaki,WosnitzaSdH}, and in some heavy fermion
superconductors \cite{Y. Inada}.  On entering the superconducting
(mixed) state from high fields, the effect of the
superconductivity is to cause an attenuation of the {\em
amplitude} of the oscillatory magnetization which can be described
by the factor $R_s$ in Eq.~\ref{DHVA_eqn}. In the normal state
$R_s(B_0)=1$.

Many theories have been developed to account for the DHVA effect
in the mixed state, all of which implicate the superconducting
energy gap, $\Delta$, in the signal attenuation \cite{JT}.   By
performing angle resolved DHVA measurements, it should be possible
to probe any anisotropy in $\Delta$, which is an issue of recent
intensive debate \cite{Schrama} in organic superconductors.
There is considerable evidence that the $\Delta$ of the
$\kappa$-type BEDT-TTF organic superconductors displays $d$-wave
symmetry with line nodes. Evidence includes NMR
\cite{DeSoto,Kanoda,Wzietek}, susceptibility \cite{Pinteric}, heat
capacity \cite{Nakazawa,wosnitzaphysicaC}, thermal conductivity
\cite{Behnia},  and tunnelling spectra \cite{Arai}, whilst heat
capacity measurements can also be understood with strong-coupling
s-wave behavior \cite{wosnitzanew}.

In this paper, we report angle resolved DHVA measurements near the
upper critical field $B_{c2}$ of
$\kappa$-(BEDT-TTF)$_{2}$Cu(NCS)$_{2}$ and at low temperatures.
Our measurements probe both the ``vortex liquid state'' and the
reduced dimensionality of the electronic properties. We observe
DHVA oscillations down to the irreversibility field.  The damping
of the DHVA signal has been investigated over a wide range of
crystal angles up to $\theta = 62^{\circ}$, and it is found that
the $R_{s}(\theta)$ curves are translated to higher fields as
$\theta$ increases, where $\theta$ is the angle between the normal
to the conducting layers and the applied magnetic field. The shape
of the curves are largely understood in terms of the component of
the applied field perpendicular to the conducting planes.

\section{Experimental Details}
The single crystals used in this experiments were grown by a
standard electrochemical method \cite{Saito}. The platelet
crystals had well developed facets, making orientation
unambiguous.  A single crystal was attached with a small amount of
vacuum grease onto the end of a flat pick-up coil, which could be
rotated in the magnetic field. Two mounting geometries were used:
in the first the crystal was rotated in the $a^{*}-b$ plane; in
the second the crystal was rotated by $\phi =45^{\circ}$ about the
$a^{*}$ axis and re-mounted on the pick-up coil. Two samples with
different size (2 mm and 1 mm in $b$-axis length) from the same
batch were examined.

Experiments were performed in a top-loading dilution refrigerator,
employing the field modulation technique to measure the DHVA
signal \cite{Shoenberg}, which was detected at the second harmonic
of the modulation frequency of 5.2 Hz.  Typical modulation
amplitudes were in the range $b$=10 mT to 25 mT.  AC
susceptibility measurements were performed at frequencies of 5.2
Hz and 93 Hz. The external field generated by a solenoid was swept
at a rate of 0.001 T/s. The temperature was measured using a
calibrated germanium thermometer in the field-compensated region
of the mixing chamber.

\section{Results}
\begin{figure}
\centering
\epsfxsize=8.0cm
\epsffile{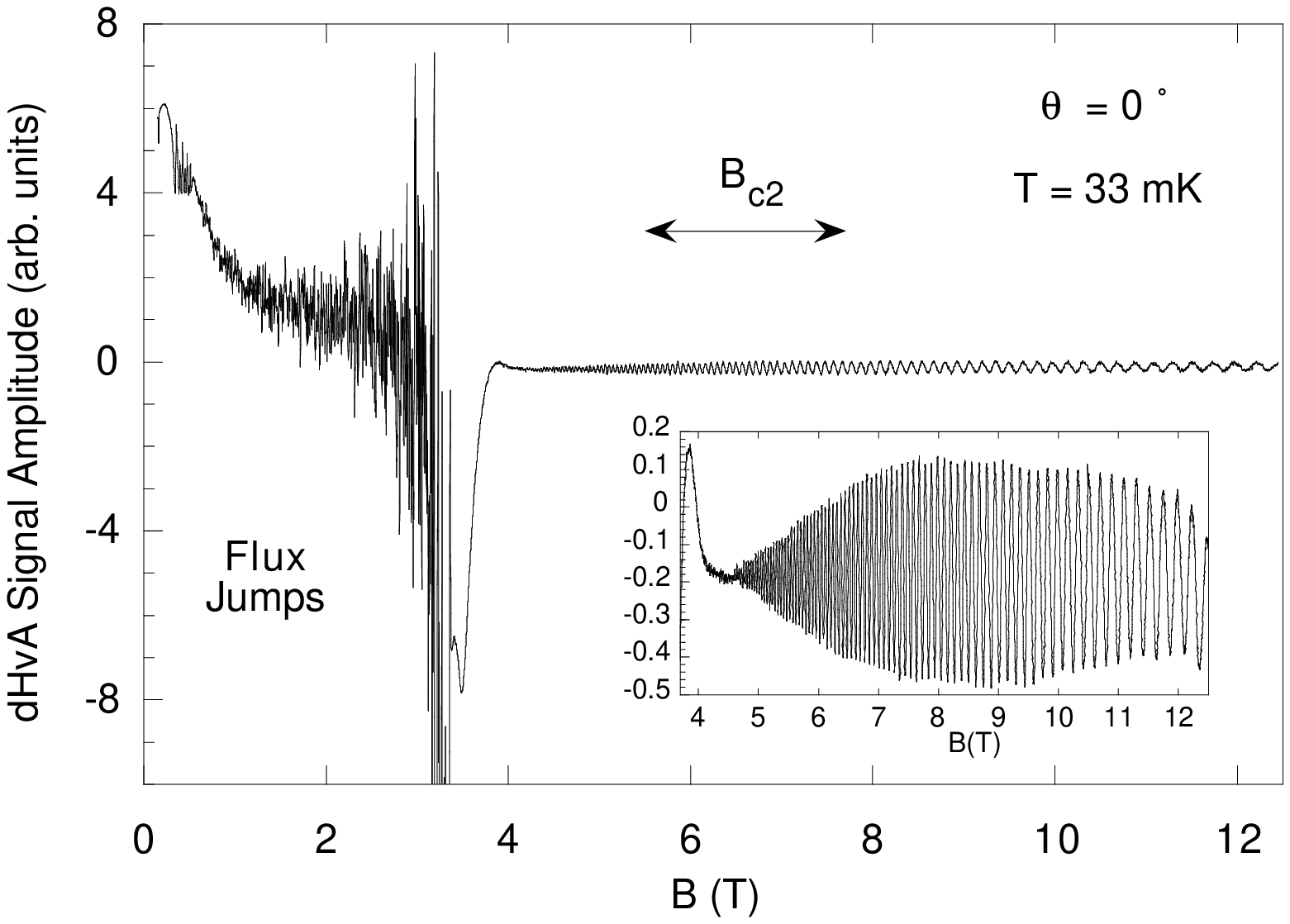}
\caption{The
oscillatory magnetization as measured by the second harmonic
pick-up, with the magnetic field applied perpendicular to the {\it
b-c} planes at a temperature of 33 mK. The DHVA oscillations are
observed above 4 Tesla with a DHVA frequency of 601 T. The noisy
signal below 3.5 T is due to flux jumps occurring within the
sample.  }
\label{Fig_DHVA}
\end{figure}

\begin{figure}
\centering
\epsfxsize=8.0cm
\epsffile{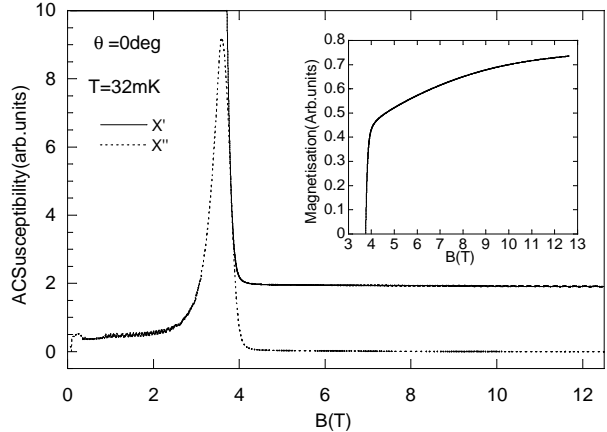}
\caption{In-phase
$\chi'$ (solid line) and quadrature $\chi''$ (dotted line)
components of the AC susceptibility of the sample in a magnetic
field applied perpendicular to the {\it b-c} plane at 32 mK.
Inset: The magnetization obtained by integrating the in-phase
component of the AC susceptibility.  The magnetization is not
linear in field above 4 T due to superconducting (diamagnetic)
fluctuations. }
\label{Fig_AC}
\end{figure}

Fig.~\ref{Fig_DHVA} shows the oscillatory magnetization measured
for T=33 mK with the magnetic field applied perpendicular to the
highly conducting {\it b-c} planes ($\theta =0^{\circ}$). DHVA
oscillations are observed above 4 T with a frequency of 601 T
(magnified in the inset), originating from the Q2D pockets of the
Fermi surface \cite{wosnitzaDHVA}. Fig.~\ref{Fig_AC} shows the
in-phase and quadrature components of the AC susceptibility
measured at 93 Hz; the inset shows the magnetisation obtained from
integrating the in-phase component with field. The feature at 3.9
T in the AC susceptibility corresponds to the irreversibility
field, $B_{\text{irr}}$, measured in the torque magnetometry
experiments of Sasaki {\it et al.} \cite{sasaki}. In this paper,
we define the irreversibility field as the field at which there is
a rapid increase in the real and imaginary parts of the
susceptibility corresponding to the onset of a strong diamagnetic
response and pinning of the vortices (see Fig.~\ref{Fig_AC}). The
relationship of the irreversibility field and the field at which
the flux lattice melts remains controversial in this
compound\cite{sasaki,Mola2}.

\begin{figure}
\centering
\epsfxsize=8.0cm
\epsffile{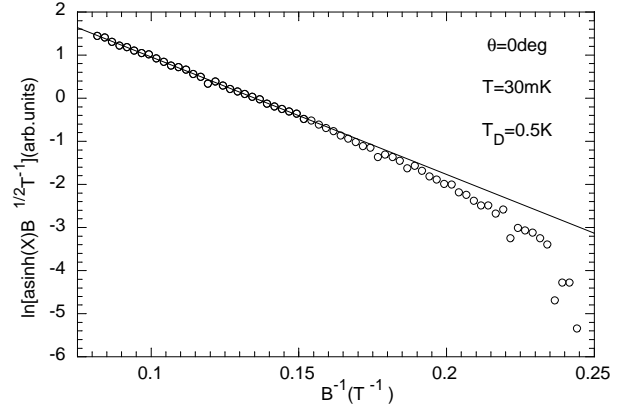}
\caption{Dingle
plot of the DHVA signal obtained with the magnetic field applied
perpendicular to the {\it b-c} plane at 30 mK.  The deviation in
linearity below 7 T represents the additional damping due to
superconductivity. $a$ is the amplitude of the DHVA oscillations,
$X$ is defined in Eq.~\protect\ref{DHVA_eqn}. }
\label{Fig_Dingle}
\end{figure}

The DHVA signal is observed only above $B_{\text{irr}}$, in the
vortex liquid phase, where the magnetization is fully reversible.
The quasiparticle effective mass has been obtained from the
temperature dependence of the DHVA signal from the Q2D pockets,
and at high fields is found to be $m^{*}=(3.5\pm 0.2)m_{e}$, in
agreement with previous studies \cite{wosnitzaDHVA}.  We found
that the 2D Lifshitz-Kosevich (L-K) expression with constant
chemical potential (Eq.~\ref{DHVA_eqn}) describes the field
dependence of the amplitude of the DHVA oscillations
\cite{Shoenberg,Harrison}. The main difference in the functional
form of the 2D and 3D L-K expression is a factor of $\sqrt{B}$. In
the case of $\kappa$-(BEDT-TTF)$_2$Cu(NCS)$_2$, the constant
chemical potential is provided by the quasi-1D sheets of the Fermi
surface, which act as a particle reservoir.  The corresponding
Dingle plot at 33 mK is shown in Fig.~\ref{Fig_Dingle}.  The plot
produced a straight line at high fields, from which a Dingle
temperature of T$_{D}=0.53\pm0.02$ K is deduced \cite{Most}, which
is comparable to other samples \cite{Ito}.  The 2D form of the L-K
expression predicts the absolute amplitude of the oscillations,
this can be compared with experimental data, by extrapolating to
infinite field ($1/B \rightarrow 0$).  By assuming the thickness
of the 2D layer in Eq.~\ref{DHVA_eqn} is that of a BEDT-TTF layer,
we obtain good agreement with Eq.~\ref{DHVA_eqn} further
validating the use of the 2D L-K expression.

\begin{figure}
\centering
\epsfxsize=8.0cm
\epsffile{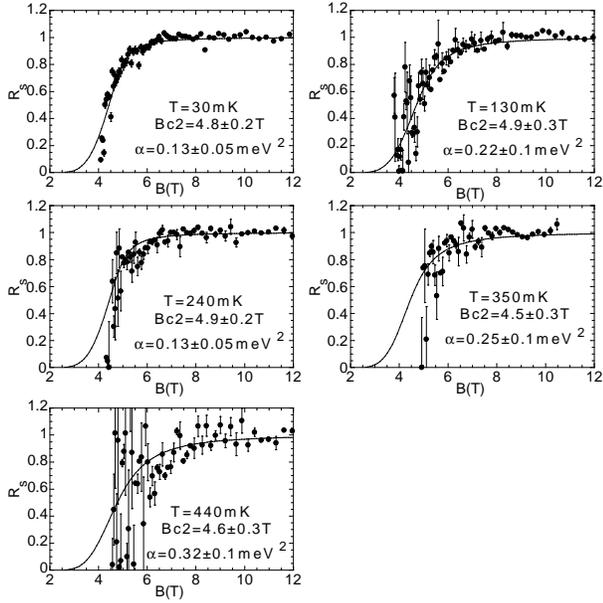}
\caption{Fits for
the observed $R_{s}$ to
Eqs.~\protect\ref{Maki_eqn1}--\protect\ref{Maki_eqn2} which
include the effects of superconducting fluctuations at $T$=30 mK,
together with the data at 130 mK, 240 mK, 350 mK, and 440 mK.  }
\label{Fig_Rs_T}
\end{figure}

The Dingle plot (Fig.~\ref{Fig_Dingle}) deviates from linearity
below $\approx$ 7 T, indicating that the DHVA signal suffers an
additional attenuation below this field, associated with the onset
of superconductivity \cite{vanderWel,Ito}.  In contrast to the
behavior in other type II superconductors, this attenuation
appears well above the $B_{\text{irr}}$. The additional
attenuation can be conveniently expressed as a field-dependent
factor amplitude, $R_{s}(B_0)$, in the L-K formula.  The
$R_s(B_0)$ values determined from the data in Fig.~\ref{Fig_DHVA}
are shown in Fig.~\ref{Fig_Rs}(a). $R_{s}$ curves have been
extracted from DHVA data taken at temperatures up to 440 mK, and
are shown in Fig.~\ref{Fig_Rs_T}. As the temperature is increased,
the signal-to-noise is steadily reduced, and this manifests itself
as increased data point scatter in the $R_{s}$ curves.  The form
of $R_{s}$ largely unchanged up to T=440 mK. The solid curves on
the plot will be discussed in the next section.

An angle resolved DHVA study has been performed with the aim of
investigating the angular dependence of $R_{s}$ and the
dimensionality of the Fermi surface. Clear DHVA oscillations are
observed at high fields, whilst at lower fields a noisy signal due
to flux jumps is observed.  The DHVA oscillations are observed
over a progressively shorter field range as $\theta$ increases,
with the position of the lower bound having a $1/\cos\theta$
angular dependence.  From the analysis of the DHVA amplitude, both
the DHVA frequency and effective mass were found to have a
$1/\cos\theta$ angular dependence.  These findings are in good
agreement with previous studies\cite{wosnitzaDHVA}, demonstrating
the 2D nature of the Fermi surface.  $R_{s}$ curves have been
extracted at each angle from the deviation of the Dingle plot
(Fig.~\ref{Fig_Dingle})with respect to the straight line.  The
attenuation due to the superconductivity occurs at higher fields
as $\theta$ increases. By considering the components of the main
and modulation magnetic fields perpendicular to the {\it b-c}
planes ($B\cos\theta$ and $b\cos\theta$ respectively), the
$R_{s}(\theta)$ curves can to be scaled to lie on top of each
other, as shown in Fig.~\ref{Fig_Rs_scale}(a), however, the
concave nature is slightly less prominent at higher angles.
$R_{s}(\theta)$ curves showed the same behavior when the crystal
was oriented such that $\phi=45^{\circ}$.

\section{Discussion}
\subsection{Additional damping of the DHVA signal in the fluctuation region}
The field dependence of the DHVA oscillations in
$\kappa$-(BEDT-TTF)$_{2}$Cu(NCS)$_{2}$ differs from other
superconductors such as NbSe$_{2}$ and V$_{3}$Si in that the
damping of the DHVA signal due to the superconductivity appears at
a field considerably greater than that where a large diamagnetic
response due to the superconductivity is observed.  Having said
this, if we integrate the AC susceptibility with field to obtain
the magnetization (see inset to Fig.~\ref{Fig_AC}), then a
deviation from the linear $M=\chi B/\mu_0$ behavior expected for a
normal metal occurs for $B \lesssim 10$~T. This deviation is
presumably due to superconducting (diamagnetic) fluctuations.  As
mentioned in the introduction, the short coherence length and
two-dimensionality of $\kappa$-(BEDT-TTF)$_{2}$Cu(NCS)$_{2}$ mean
that fluctuations in the superconducting order parameter will be
large over a wide temperature and field range.  Thus there is no
`sharp' transition into the mixed state at $B_{c2}$.  Rather,
$B_{c2}$ should be thought of as the cross-over field associated
with the formation of a flux liquid phase.  It has been
argued\cite{Moore1997,Forgan}that is this picture is actually
appropriate for all type-II superconductors.  The cross over
region being smaller in 3D systems.

Understanding the DHVA effect in the superconducting mixed state
is a challenging problem from the theoretical viewpoint.  The
experimental observation of these oscillations has stimulated many
theoretical studies over the last ten years \cite{JT}.  A feature
common to many of the theories is the prediction of electronic
states within the energy gap, which appear in a plane
perpendicular to the applied magnetic field
\cite{Brandt,Maki,Miyake,MG,DT,NMA,Gorkov}; it is the presence of
these states which give rise to the DHVA oscillations in the mixed
state.  In $\kappa$-(BEDT-TTF)$_{2}$Cu(NCS)$_{2}$, the presence of
a quasiparticle states is supported by thermal
conductivity\cite{Behnia} and heat capacity\cite{Nakazawa}
measurements in a magnetic field . Most of the theories ascribe
the additional damping of the DHVA signal to the additional
scattering of the quasiparticles by the vortex lattice. In the
Maki, Stephen approach, the vortex lattice is treated as a random
media to scatter the quasiparticles \cite{Maki,Stephen,Wasserman}.
Dukan and Tesanovic \cite{DT}, Norman, MacDonald, and Akera
\cite{NMA}, Maniv {\it et al.} \cite{Maniv}, Zhuravlev {\it et
al.} \cite{Zhuravlev},Gvozdikov and Gvozdikova \cite{GG}, treat
the structure of the vortex lattice in detail.

\begin{figure}
\centering
\epsfxsize=8.0cm
\epsffile{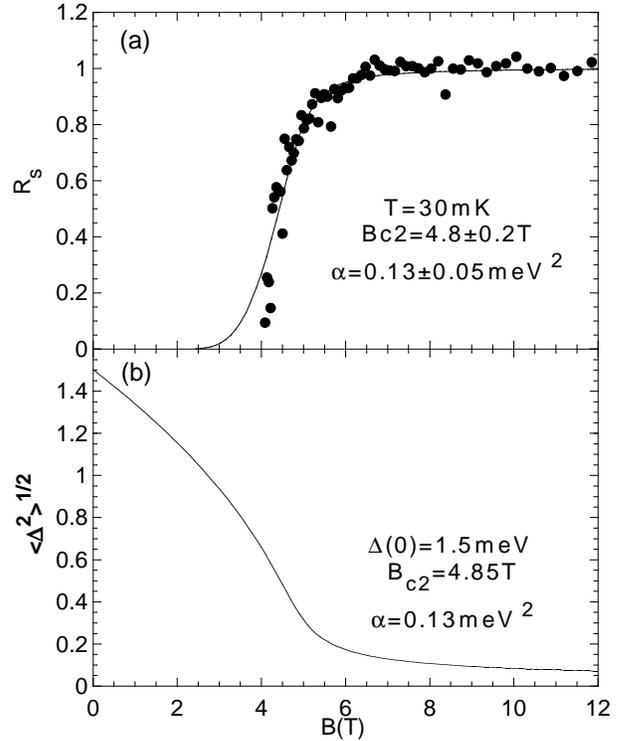}
\caption{ (a) Damping of the DHVA signal due to superconductivity,
$R_{s}$, at 30 mK in a magnetic field applied perpendicular to the
{\it b-c} plane ($\theta=0^{\circ}$).  The error bars are caused
by the experimental uncertainty in the normal state scattering
rate. The solid line is a fit of
Eqs.~\protect\ref{Maki_eqn1}--\protect\ref{Maki_eqn2}. (b) The
root-mean-square order parameter $\sqrt{\left< \Delta^2 \right>}$
as given by Eq.~\protect\ref{Maki_eqn1}.}
\label{Fig_Rs}
\end{figure}

Attempts to model the experimental form of the $R_{s}$ curves
using the available theories describing the DHVA effect in the
mixed state, coupled with the mean field form for the
field-dependence of the superconducting gap, were unsuccessful in
$\kappa$-(BEDT-TTF)$_{2}$Cu(NCS)$_{2}$ . The existing theories do
not reproduce the downward curvature in $R_s$ observed above about
4.5 T in Fig.~\ref{Fig_Rs}. A successful explanation of the
$R_{s}$ curves may need to account for the effects of the
superconducting fluctuations inherent in this
material\cite{friemelfluc,ito,lang,Behnia,Graebner}. One approach
is to assume that the DHVA effect is sensitive to the square (or
root mean square) of the order parameter $\left< \Delta^{2}
\right>$.  If the fluctuating part of the order parameter is
$\delta \Delta$, then $\left< \Delta^{2} \right> = \left< \Delta
\right>^2 + \left< (\delta \Delta)^{2} \right>$. An approximate
form\cite{Makip,Ito} for the fluctuating order parameter can be
obtained using the lowest Landau level approximation $\left<
\delta
\Delta^{2} \right> \propto 1/(B/B_{c2}-1)$. Far below
$B_{c2}$, the superconducting gap takes a mean-field form, $\Delta
(B)=\Delta(0)\sqrt{1-B/B_{c2}}$. We can interpolate between the
two limiting regimes with the formula,
\begin{eqnarray}
\label{Maki_eqn1} \langle \Delta^{2}\rangle & = &
\sqrt{\left[\frac{\Delta(0)^{2}}{2}\left(1-\frac{B}{B_{c2}}\right)\right]^{2}
+\alpha(T)^{2}} \nonumber \\ & &
+\frac{\Delta(0)^{2}}{2}\left(1-\frac{B}{B_{c2}}\right).
\end{eqnarray}
The parameter $\alpha = \left< \Delta^{2} \right>_{B_{c2}}$
determines the strength of the fluctuations at $B_{c2}$, where
$\left< \Delta \right> = 0$. Since the DHVA oscillations are
always observed in the vortex liquid state above $B_{\text{irr}}$,
we employ the Maki, Stephen approach in which the vortices are
treated as a random media which scatter the quasiparticles
\cite{Maki,Stephen,Wasserman},
\begin{eqnarray}
\label{Maki_eqn2}
R_{s} = \exp \left[ -\pi^{3/2} \langle \Delta^{2}\rangle \left(
\frac{m_{b}}{\hbar e B} \right)^{2} \left( \frac{B}{F} \right)^{1/2} \right].
\end{eqnarray}
The interpolation formula allows the data taken at $T$=30 mK to be
fitted with $B_{c2}=4.8\pm 0.2$ T and $\alpha =0.13\pm 0.05$
meV$^2$ as fitting parameters \cite{Ito,halffactor}.  In the
fitting procedure, the following parameters are used. For the
superconducting gap at zero magnetic field, a value $\Delta
(0)=1.5$ meV determined from the BCS relation
$2\Delta=3.5k_{B}T_{c}$ with $T_{c}$=9 K. For the band mass, a
value $m_{b}=1.2m_{e}$ found by cyclotron resonance measurement is
used \cite{bandmass}.  We obtain a somewhat smaller $B_{c2}$ value
compared with $B_{c2} \sim 6$T that is given by Sasaki {\it et
al}, based on the magnetic field where torque
measurements\cite{sasaki} of the DHVA amplitude start to deviate
from the prediction of the normal L-K expression by their torque
measurement. In our analysis, we allow the deviation from the
normal state expression even above $B_{c2}$ [see
Fig.~\ref{Fig_Rs}(b)], since the effects of fluctuations are taken
into account.

Fig.~\ref{Fig_Rs_T} shows $R_{s}$ curves extracted at temperatures
up to T=440 mK. The $B_{c2}$ values given by the fluctuation
analysis at each temperature are almost unchanged and are in the
range of 4.5--5 T, in good agreement with the weak temperature
dependence of the $B_{c2}$ in this temperature region
\cite{sasaki}.

\begin{figure}
\centering
\epsfxsize=8.0cm
\epsffile{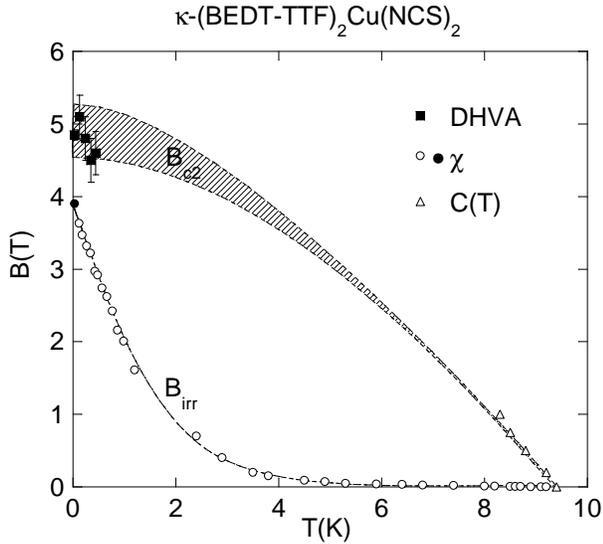}
\caption{Superconducting phase diagram for $\kappa$-
(BEDT-TTF)$_{2}$Cu(NCS)$_{2}$ : squares are $B_{c2}$ determined by
fitting Eqs.~\protect\ref{Maki_eqn1}--\protect\ref{Maki_eqn2} to
DHVA data; triangles are $B_{c2}$ determined from the specific
heat anomaly (Ref.~\protect\onlinecite{Graebner}). The
irreversibility field $B_{\text{irr}}$ is determined from AC
susceptibility measurements: closed circles (this work); open
circles (Ref.~\protect\onlinecite{sasaki})  }
\label{Fig_phase}
\end{figure}

Fig.~\ref{Fig_phase} summarizes the superconducting phase diagram
resulting from the present work.  $B_{\text{irr}}$ is obtained
from the AC susceptibility \cite{sasaki} and the present work.
$B_{c2}$ is determined from the DHVA effect via
Eq.~\ref{Maki_eqn1}--\ref{Maki_eqn2}.  Because $B_{c2}$ is treated
as a ``crossover field'' defined through Eq.~\ref{Maki_eqn1}, the
values are different from others reported in the literature
\cite{sasaki,lang}.

\subsection{The dimensionality of the superconductivity of the
$\kappa$-(BEDT-TTF)$_{2}$Cu(NCS)$_{2}$}
\begin{figure}
\centering
\epsfxsize=8.0cm
\epsffile{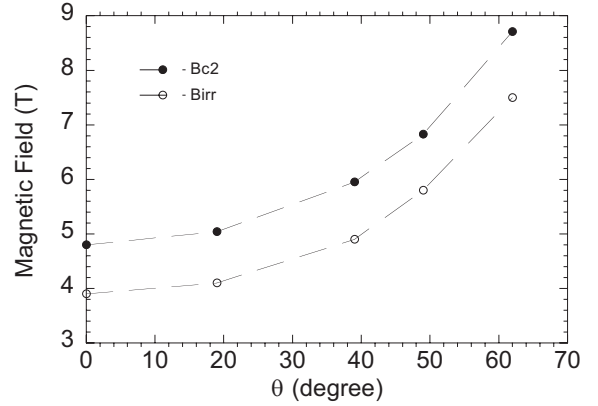}
\caption{The
angular dependence of $B_{c2}$ and $B_{\text{irr}}$ determined
from DHVA and AC susceptibility measurements respectively.  Lines
are fits to a $1/\cos\theta$ dependence.}
\label{Fig_Bc2_angle}
\end{figure}

We now discuss the angle
dependence of the DHVA signal.  The anomaly in the AC
susceptibility curve representing the irreversibility field
$B_{\text{irr}}$ and $B_{c2}$ can be approximately scaled with
$B\cos\theta$ as shown in Fig.~\ref{Fig_Bc2_angle}. The
$B_{\text{irr}}$ was found to have a 1/$\cos\theta$ angular
dependence, in agreement with the angle-dependent torque
measurements \cite{Kawamata,Mola2}. This may be taken as evidence
for strong 2D nature of the superconductivity in
$\kappa$-(BEDT-TTF)$_{2}$Cu(NCS)$_{2}$.

\begin{figure}
\centering
\epsfxsize=8.0cm
\epsffile{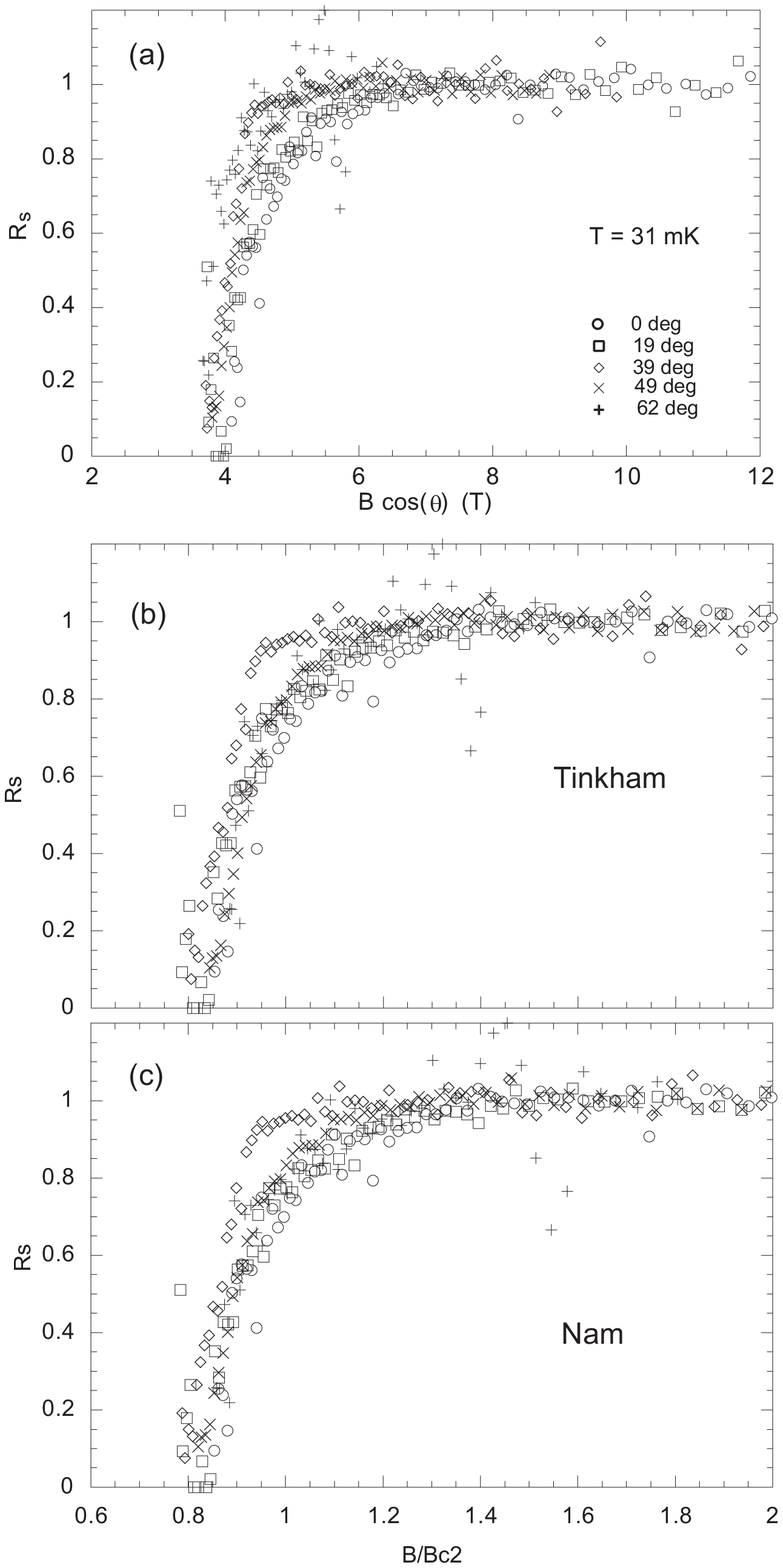}
\caption{ Scaling
of $R_{s}(\theta)$: (a) assuming $R_s$ depends only on
$B\cos\theta$; (b) assuming $R_s$ depends only on $B_{c2}$ with
$B_{c2}$ given by Tinkham\protect\cite{Tinkham}; (c) assuming
$R_s$ depends only on $B_{c2}$ with $B_{c2}$ given by Nam et
al.\protect\cite{Nam}. }
\label{Fig_Rs_scale}
\end{figure}

In Fig.~\ref{Fig_Rs_scale}(a), the $R_{s}(\theta)$ curves at each
angle of the magnetic field are shown scaled onto one another by
taking the component of the magnetic field perpendicular to the
layers $B\cos\theta$. The behavior of $R_{s}$ curves at different
angles $\theta$ scales approximately with $B\cos\theta$. Thus the
field perpendicular to the conducting layers appears to control
the superconducting properties of
$\kappa$-(BEDT-TTF)$_{2}$Cu(NCS)$_{2}$. However, in detail, the
behavior of $R_{s}$ curves at different angles $\theta$ is not
fully scaled with $B\cos\theta$ near $B_{c2}$.  The concave nature
of $R_{s}$ at the $B_{c2}$ seems to be less prominent at the
higher angle region.  At $\theta=49^{\circ}$ and 62$^{\circ}$,
appropriate value for $\alpha =0.1$ meV$^2$ is slightly smaller
than that at lower angles.  Since we have larger ambiguity in
Dingle analysis at higher angles, apparent angle dependence in
$R_{s}$ may be an artifact.  For a quasi-2D superconductor with
sufficiently weak coupling between layers, Tinkham proposed an
angular dependence of the upper critical field expressed as
$|B_{c2}\cos\theta/B_{c2}^{\bot}|+(B_{c2}\sin\theta/B_{c2}^{\parallel})^{2}=1$,
where the $B_{c2}^{\bot}$ and $B_{c2}^{\parallel}$ are the upper
critical field in the perpendicular and parallel to the 2D layer
\cite{Tinkham}.  If we rescale the magnetic field $B/B_{c2}$ with
$B_{c2}^{\parallel}=20$ T in Tinkham's model, all $R_{s}$ data
except one with $\theta=39^{\circ}$ can be scaled onto one another
as shown in Fig.~\ref{Fig_Rs_scale}(b).  Recently, Nam {\it et al}
proposed a new scaling form for the angle dependence of the
$B_{c2}$ \cite{Namnew}.  The $B_{c2}$ of
$\kappa$-(BEDT-TTF)$_2$Cu(NCS)$_2$ would be comprised of orbital
and spin contributions.  Their scaling form is
$B_{c2}(\theta)=B_0/\sqrt{\cos^2\theta+\alpha^2}$, where
$B_0=(1+\alpha^2)B_{c2}(\theta=0)$ and
$\alpha=B_0/B_{\text{spin}}$. This expression also fits the angle
dependence of $R_{s}$ with $B_{\text{spin}}=30$ T, as shown in
Fig.~\ref{Fig_Rs_scale}(c).

\section{Conclusion}
We have observed the effects of fluctuations and reduced
dimensionality on the de Haas-van Alphen (DHVA) signal from
$\kappa$-(BEDT-TTF)$_{2}$Cu(NCS)$_{2}$. The DHVA signal suffers an
additional attenuation in the mixed state.  The attenuation is
analyzed in terms of the superconducting fluctuation effects. Such
an analysis can be used to determine the superconducting phase
diagram of $\kappa$-(BEDT-TTF)$_{2}$Cu(NCS)$_{2}$. In particular,
yielding $B_{c2}$, which we interpret as a cross over field
associated with the formation of the vortex flux lattice. The
angular dependence of this damping has been measured over a wide
range of angles. Our results scale well with the component of
applied field perpendicular to the conducting planes suggesting
that the superconductivity is highly two dimensional in this
material.

\section*{Acknowledgments}
We thank Prof. K. Maki and Dr N. Harrison for helpful discussions.
One of the authors, H. I., acknowledges the support from the
overseas research fellowship from the Ministry of Education,
Science, Sports, and Culture of Japan.

\begin{references}

\bibitem{Saint} T. Ishiguro, K. Yamaji and G. Saint, {\it Organic
Superconductors, 2nd ed.,} (Springer Verlag, Berlin 1998).

\bibitem{Oshima} K. Oshima, T. Mori, H. Inokuchi, H. Urayama, H. Yamochi and
G. Saint, Phys. Rev. B {\bf 38}, 938 (1988).

\bibitem{ito93} H. Ito, T. Ishiguro, T. Komatsu, N. Matsukawa, G. Saint, and
H. Anzai, J. Supercond. {\bf 7}, 667 (1994).

\bibitem{friemelfluc} S. Friemel, C. Pasquier, Y. Loirat, and D. Jerome,
Physica C {\bf 259}, 181 (1996).

\bibitem{friemelangle} S. Friemel, C. Pasquier and D. J\'{e}rome, J. Phys. I
France {\bf 6}, 2043 (1996).

\bibitem{brooks} F. Zuo, J.S. Brooks, R.H. McKenzie, J.A. Schlueter and
J.M. Williams, Phys. Rev. B {\bf 61}, 750 (2000).

\bibitem{ito} H. Ito, Y. Nogami, T. Komatsu, G. Saint, and N. Hosoito, JJAP
Series 7, {\it Mechanism of Superconductivity}, 419 (1992).

\bibitem{lang} M. Lang, F. Steglich, N. Toyota, and T. Sasaki, Phys. Rev. B
{\bf 49}, 15227 (1994).

\bibitem{Kanoda1271} K. Kanoda, K. Akiba, K. Suzuki, T. Takahashi, and
G. Saint, Phys. Rev. Lett. {\bf 65}, 1271 (1990).

\bibitem{Mansky} P. A. Mansky, P. M. Chaikin, and R. C. Haddon, Phys. Rev. B
{\bf 50}, 15929 (1994).

\bibitem{Farrell} D. E. Farrell, C. J. Allen, R. C. Haddon, and
S. V. Chichester, Phys. Rev. B {\bf 42}, 8694 (1990).

\bibitem{Kawamata} S. Kawamata, K. Okuda, T. Sasaki, and N. Toyota, Solid
State Commun. {\bf 89}, 955 (1994).

\bibitem{sasaki} T. Sasaki, W. Biberacher, K. Neumaier, W. Hehn, K. Andres,
and T. Fukase, Phys. Rev. B {\bf 57}, 10889 (1998).

\bibitem{Mola2} M. M. Mola, S. Hill, J. S. Brooks and
J. S. Qualls,Phys. Rev. Lett. {\bf 86}, 2133 (2001).

\bibitem{Graebner} J.E. Graebner, R.C. Haddon, S.V. Chichester and
S. H. Glarum, Phys. Rev. B {\bf 41}, 4808 (1990).

\bibitem{Behnia} S. Belin, K. Behnia, and A. Deluzet, Phys. Rev. Lett. {\bf
81}, 4728 (1998).

\bibitem{Mota} A. C. Mota, Physica C {\bf 185-189}, 343 (1991).

\bibitem{Nishizaki} T. Nishizaki, T. Sasaki, T. Fukase, and N. Kobayashi,
Phys. Rev. B {\bf 54}, R3760 (1996).

\bibitem{Pasquire} C. Pasquire and S. Friemel, Synth. Met. {\bf 103},1845
(1999).

\bibitem{Shoenberg} D. Shoenberg, {\it Magnetic oscillations in metals}
(Cambridge University Press, Cambridge, 1984).

\bibitem{Harrison} N. Harrison, R. Bogaerts, P. H. P. Reinders, J. Singleton,
S. J. Blundell, and F. Herlach, Phys.  Rev. B {\bf 54} (1996)
9977.

\bibitem{Champel2001} T. Champel and V. P. Mineev, Phil. Mag. B, {\bf 81}, 55
(2001).

\bibitem{GR} J. E. Graebner and M. Robbins, Phys. Rev. Lett. {\bf 36},
422(1976).

\bibitem{JT} T. J. B. M. Janssen, C. Haworth, S. M. Hayden, P. Meeson and
M. Springfield, Phys. Rev. B {\bf 57}, 11698 (1998), and references
therein.

\bibitem{Terashima} T. Terashima, C. Haworth, H. Takeya, S. Uji and H. Aoki,
Phys. Rev. B {\bf 56}, 5120 (1997).

\bibitem{vanderWel} P. J. van der Wel, J. Caulfield, R. Corcoran, P. Day,
S. M. Hayden, W. Hayes, M. Kurmoo, P. Meeson, J. Singleton, and M.
Springfield, Physica C {\bf 235-240}, 2453 (1994); P. J. Van der
Wel, S. M. Hayden, M. Springfield, P. Meeson, J. Caulfield, J.
Singleton, W. Hayes, M. Kurmoo and P. Day, Synth. Met. {\bf 70},
831 (1995).

\bibitem{WosnitzaSdH} J. Wosnitza, S. Wanka, J. Hagel, R. H\"{a}ussler,
H.V. L\"{o}hneysen, J. A. Schlueter, U. Geiser, P. G. Nixon, R. W.
Winter, andG. L. Gard, Phys. Rev. B {\bf 62}, R11973 (2000).

\bibitem{Y. Inada} Y. Inada and Y. Onuki, Low Temp. Phys. {\bf 25}, 573
(1999), and references therein.

\bibitem{Schrama} J. M. Schrama, E. Rzepniewski, E. S. Edwards, J. Singleton,
A. Ardavan, M. Kurmoo, and P.  Day, Phys. Rev. Lett. {\bf 83},
3041 (1999), S. Hill, N. Harrison, M. Mola, and J. Wosnitza, Phys.
Rev. Lett. {\bf 86}, 3451 (2001).

\bibitem{DeSoto} S. M. De Soto, C. P. Slichter, A. M. Kini, H. H. Wang,
U. Geiser, and J. M. Williams, Phys.  Rev. B {\bf 52}, 10364
(1995).

\bibitem{Kanoda} K. Kanoda, K. Miyagawa, A. Kawamoto, and Y. Nakazawa,
Phys. Rev. B {\bf 54}, 76 (1996).

\bibitem{Wzietek} P. Wzietek, H. Mayaffre, D. Jerome and S. Brazovskii,
J. Phys. I France {\bf 6}, 2011 (1996), and references therein.

\bibitem{Pinteric} M. Pinteri\'{c}, S. Tomi\'{c}, M. Prester, D. Drobac,
O.Milat, K. Maki, D. Schweitzer, I.  Heinen, and W. Strunz, Phys.
Rev. B {\bf61}, 7033 (2000), and references therein.

\bibitem{Nakazawa} Y. Nakazawa and K. Kanoda, Phys. Rev. B {\bf 55} R8670
(1997); Y. Nakazawa and K.  Kanoda, Synth. Met. {\bf 85}, 1563
(1997).

\bibitem{wosnitzaphysicaC} J. Wosnitza, Physica C {\bf 317}, 98 (1999).

\bibitem{Arai}T. Arai, K. Ichimura, K. Nomura, S. Takasaki, J. Yamada,
S. Nakatsuji, and H. Anzai, Phys Rev.  B {\bf 63}, 104518 (2001).

\bibitem{wosnitzanew} H. Elsinger, J. Wosnitza, S. Wamka, J. Hagel,
D. Schweitzer, and W. Strunz, Phys. Rev.  Lett. {\bf 84}, 6098
(2000).

\bibitem{wosnitzaDHVA} J. Wosnitza, G. W. Crabtree, H. H. Wang, U. Geiser,
J. M. Williams and K. D. Carlson, Phys. Rev. B {\bf 45} 3018
(1992).

\bibitem{Most} The Dingle temperature is obtained using the
effective mass determined from the temperature dependence of the
DHVA signal, as has been done in this work.  Strictly speaking,
the band mass should be used.  The Dingle temperature obtained
using the band mass is 1.4 K.

\bibitem{Ito} H.Ito, S.M. Hayden, P.J. Meeson, M. Springfield and G. Saint,
J. Supercond. {\bf 12}, 525 (1999).

\bibitem{Moore1997} M. A. Moore, Phys. Rev. B {\bf 55}, 14136 (1997).

\bibitem{Forgan} E.M. Forgan, J. Phys: Cond. Matt. {\bf 11}, 7685 (1999).

\bibitem{Brandt} U. Brandt, W. Pesch, and L. Tewordt, Z. Phys. {\bf 201},209
(1967).

\bibitem{Maki} K. Maki, Phys. Rev. B {\bf 44}, 2861 (1991).

\bibitem{Miyake} K. Miyake, Physica B {\bf 186-188}, 115 (1993)

\bibitem{MG} P. Miiler and B. L. Gy\"{o}rffy, J. Phys.: Cond. Matt. {\bf 7},
5579 (1995).

\bibitem{DT} S. Dukan and Z. Te\'{s}anovi\'{c}, Phys. Rev. Lett. {\bf 74} 2311
(1995).

\bibitem{NMA} M. R. Norman, A. H. MacDonald, and H. Akera, Phys. REv. B {\bf
51}, 5927 (1995); M. R.  Norman and A. H. MacDonald, Phys. Rev. B
{\bf 54}, 4239 (1996).

\bibitem{Gorkov} L. P. Gor'kov, Pi'sma Zh. \'{E}ksp. Ther. Fiz. {\bf 68},705
(1998) [JETP Lett. {\bf 68}, 738 (1998)].

\bibitem{Stephen} M. J. Stephen, Phys. Rev. B {\bf 43}, 1212 (1991); {\it
ibid}, {\bf 45},5481 (1992).

\bibitem{Wasserman} A. Wasserman and M. Springfield, Physica C {\bf 194-196},
1801 (1994).

\bibitem{Maniv} T. Maniv, A. Y. Rom, I. D. Vagner, and P. Wyder, Solid State
Commun. {\bf 101}, 621 (1997).

\bibitem{Zhuravlev} V. N. Zhuravlev, T. Maniv, I. D. Vagner, and P. Wyder,
Phys. Rev. B {\bf 56}, 14693 (1997).

\bibitem{GG} V. N. Gvozdikov and M. V. Gvozdikova, Phys. Rev. B {\bf 58}, 8716
(1998).

\bibitem{Makip} K. Maki, private communication.

\bibitem{halffactor} According to Maki, a factor of 1/2 is mutiplied by the
$\Delta(0)^{2}$ as an average factor of the phase of the pair
potential in the case of the d-wave superconductor.  This
numerical factor will raise the appropriate $\alpha$ value up to
0.2 meV$^2$ with which the data is explained with the same
$\Delta(0)=1.5$ meV.

\bibitem{bandmass} S. Hill, J. Singleton, F. L. Pratt, M. Doporto, W. Hayes,
T. J. B. M. Janssen, A. J.  Perenboom, M. Kurmoo and P. Day,
Synth. Met. {\bf 56}, 2566 (1993).

\bibitem{Tinkham} M. Tinkham, {\it Introduction to Superconductivity, 2nd
ed.,} (McGraw-Hill, International Editions 1996).

\bibitem{Namnew} M-S Nam, J. A. Symington, J. Singleton, S. J. Blundell,
A. Ardavan, J. A. A. J. Perenboom, M. Kurmoo, and P. Day, J.
Phys.: Cond. Matt. {\bf 11} (1999) L477.

\bibitem{Nam} M.-S. Nam, M.M. Honald, C. Proust, N. Harrison, C.H. Mielke,
S.J. Blundell, J. Singleton, W.  Hayes, M. Kurmoo and P. Day,
Synth. Met. {\bf 103}, 1905 (1999).

\end{references}
\end{document}